\documentstyle[preprint,aps,eqsecnum,epsf,floats]{revtex}
\begin{document}
\tighten

\def\bfl{{\bbox \ell}}

\newcommand{\gsim}{\raisebox{-0.7ex}{$\stackrel{\textstyle >}{\sim}$ }}
\newcommand{\lsim}{\raisebox{-0.7ex}{$\stackrel{\textstyle <}{\sim}$ }}

\def\bull{\vrule height .9ex width .8ex depth -.1ex}
\def\MeV{{\rm MeV}}
\def\GeV{{\rm GeV}}
\def\Tr{{\rm Tr\,}}
\def\nrcpt{NR\raise.4ex\hbox{$\chi$}PT\ }
\def\ket#1{\vert#1\rangle}
\def\bra#1{\langle#1\vert}
\def\ltap{\ \raise.3ex\hbox{$<$\kern-.75em\lower1ex\hbox{$\sim$}}\ }
\def\gtap{\ \raise.3ex\hbox{$>$\kern-.75em\lower1ex\hbox{$\sim$}}\ }
\def\abs#1{\left| #1\right|}   
\def\CA{{\cal A}}
\def\CC{{\cal C}} 
\def\CD{{\cal D}}
\def\CE{{\cal E}}
\def\CL{{\cal L}}
\def\CO{{\cal O}}
\def\CZ{{\cal Z}}
\def\bvert{\Bigl\vert\Bigr.}
\def\pds{{\it PDS}\ }
\def\ms{MS}
\def\ddq{{{\rm d}^dq \over (2\pi)^d}\,}  
\def\ddqm{{{\rm d}^{d-1}{\bf q} \over (2\pi)^{d-1}}\,}
\def\bfq{{\bf q}}
\def\bfk{{\bf k}}
\def\bfp{{\bf p}}
\def\bfpp{{\bf p '}}
\def\bfr{{\bf r}}
\def\dtr{{\rm d}^3\bfr\,}
\def\bfx{{\bf x}}
\def\dtx{{\rm d}^3\bfx\,}
\def\dfx{{\rm d}^4 x\,}
\def\bfy{{\bf y}}
\def\dty{{\rm d}^3\bfy\,}
\def\dfy{{\rm d}^4 y\,}
\def\dfq{{{\rm d}^4 q\over (2\pi)^4}\,}
\def\dfk{{{\rm d}^4 k\over (2\pi)^4}\,}
\def\dfl{{{\rm d}^4 \ell\over (2\pi)^4}\,}
\def\dtq{{{\rm d}^3 {\bf q}\over (2\pi)^3}\,}
\def\dtk{{{\rm d}^3 {\bf k}\over (2\pi)^3}\,}
\def\dtl{{{\rm d}^3 {\bfl}\over (2\pi)^3}\,}
\def\dt{{\rm d}t\,}
\def\frac#1#2{{\textstyle{#1\over#2}}}
\def\darr#1{\raise1.5ex\hbox{$\leftrightarrow$}\mkern-16.5mu #1}
\def\){\right)}  
\def\({\left( }
\def\]{\right] } 
\def\[{\left[ }
\def\si{{}^1\kern-.14em S_0}
\def\siii{{}^3\kern-.14em S_1}
\def\diii{{}^3\kern-.14em D_1}
\def\dtwiii{{}^3\kern-.14em D_2}
\def\dthiii{{}^3\kern-.14em D_3}
\def\pziii{{}^3\kern-.14em P_0}
\def\poiii{{}^3\kern-.14em P_1}
\def\ptiii{{}^3\kern-.14em P_2}
\def\ipi{{}^1\kern-.14em P_1}
\def\idii{{}^1\kern-.14em D_2}
\def\fm{{\rm\ fm}} 
\def\MeV{{\rm\ MeV}}
\def\CA{{\cal A}}
\def\Czzm{ {\cal A}_{-1[00]} }
\def\Cttm{{\cal A}_{-1[22]} }
\def\Ctzm{{\cal A}_{-1[20]} }
\def\Cztm{ {\cal A}_{-1[02]} }
\def\Czzz{{\cal A}_{0[00]} }
\def\Cttz{ {\cal A}_{0[22]} } 
\def\Ctzz{{\cal A}_{0[20]} }  
\def\Cztz{{\cal A}_{0[02]} }

\def\Ames{ A }  

\newcommand{\eqn}[1]{\label{eq:#1}}
\newcommand{\refeq}[1]{(\ref{eq:#1})}
\newcommand{\eq}{eq.~\refeq}
\newcommand{\eqs}[2]{eqs.~(\ref{eq:#1}-\ref{eq:#2})}
\newcommand{\eqsii}[2]{eqs.~(\ref{eq:#1}, \ref{eq:#2})}
\newcommand{\Eq}{Eq.~\refeq} 
\newcommand{\Eqs}{Eqs.~\refeq}

\def\Journal#1#2#3#4{{#1} {\bf #2}, #3 (#4)}

\def\NCA{\em Nuovo Cimento}
\def\NIM{\em Nucl. Instrum. Methods}   
\def\NIMA{{\em Nucl. Instrum. Methods} A}
\def\NPB{{\em Nucl. Phys.} B}
\def\NPA{{\em Nucl. Phys.} A}
\def\PLB{{\em Phys. Lett.} B}
\def\PRL{\em Phys. Rev. Lett.}
\def\PRD{{\em Phys. Rev.} D}
\def\PRC{{\em Phys. Rev.} C}
\def\PRA{{\em Phys. Rev.} A} 
\def\ZPC{{\em Z. Phys.} C}
\def\SJP{{\em Sov. Phys. JETP}}

\def\FBS{{\em Few Body Systems Suppl.}}
\def\IJMP{{\em Int. J. Mod. Phys.} A}
\def\UJP{{\em Ukr. J. of Phys.}}


\def\spol{\alpha_{E0}}
\def\qpol{\alpha_{E2}}
\def\Mspol{\beta_{M0}}
\def\Mqpol{\beta_{M2}}

\preprint{\vbox{
\hbox{ NT@UW-98-33}
}}
\bigskip
\bigskip
\bigskip
\bigskip

\title{Tensor Polarized $\gamma$-Deuteron Compton Scattering in Effective
Field Theory}
\author{Jiunn-Wei Chen\footnote{\tt jwchen@phys.washington.edu}}
\address{Department of Physics, University of Washington, Seattle,
WA 98195-1560, USA }
\maketitle

\begin{abstract}
The differential cross section for
$\gamma$-deuteron Compton scattering from a tensor polarized deuteron is
computed in an
effective field theory. The first
non-vanishing contributions to this differential cross section are the
interference terms
between the leading electric coupling diagrams and the subleading single
potential
pion exchange diagrams or the subleading magnetic moment coupling
diagrams. 
At 90$^{\circ }$
photon scattering angle, only the pion term 
contributes at this order to the tensor polarized differential cross
section. This
provides a clean way to study the photon pion dynamics in the two nucleon
sector. The effect is measurable for photon
energies between 40 and 80 MeV provided the uncertainty in the
measured cross sections are 
$\lsim 7\%$.
\end{abstract}
\vskip 1in

\vfill\eject


\section{Introduction}

$\gamma $-deuteron Compton scattering probes the
structure of the deuteron and provides necessary information that may
allow the extraction of
neutron properties, such as neutron electric and magnetic
polarizabilities. The
differential cross section of unpolarized $\gamma $-deuteron Compton
scattering has been measured at incident photon energies of 49 MeV and 69
MeV with 10\% uncertainty \cite{Lucas}. Theoretical calculations based
on
potential models and taking the nucleon polarizabilities as inputs find
reasonably good agreement with data \cite{LLa,WWA,KarMil} but are not
sufficient
to give tight constraints on the nucleon polarizabilities. 

In comparison to potential model calculations, a model independent,
parameter free and analytic computation of unpolarized $\gamma $-deuteron
Compton scattering based on a recently developed nucleon-nucleon effective
field theory \cite
{Weinberg1,KoMany,Parka,KSWa,CoKoM,DBK,cohena,Fria,Sa96,LMa,GPLa,Adhik,RBMa,Bvk,aleph,Parkb,Gegelia,steelea,KSW,KSW2,CHa,MeStew,EGKM}
was presented in \cite{gamd}. Contributions up to next-to-leading order
(NLO) in the effective field theory expansion including diagrams
contributing to the nucleon polarizabilities give excellent agreement with
the data at 49 MeV and reasonable agreement with the data at 69 MeV. At this
order (NLO), the agreement with the data is comparable to the agreement
between the data and potential model calculations \cite{LLa,WWA}, while
inclusion of the next-next-to-leading order (NNLO) contributions should
reduce the theoretical uncertainty from 10\% to a few percent
level.

In addition to $\gamma $-deuteron Compton scattering, the effective field
theory technique also successfully describes the $NN$ scattering phase
shifts up to center-of-mass momenta of ${\bf p}\sim 300\ {\rm MeV}$ per
nucleon \cite{KSW} in all partial waves. The electromagnetic moments, form
factors~\cite{KSW2} and polarizability~\cite{dpol} of the deuteron as well
as parity violation in the two-nucleon sector~\cite{PVeft} also have been
explored with this new effective field theory. The results all agree with
data (where available) within the uncertainty from neglecting higher
order effects.

In the case of polarized $\gamma $-deuteron Compton scattering, no
experiments have been performed so far. Planned experiments at TUNL to
examine the Gerasimov-Drell-Hearn (GDH) sum rule
using a circularly polarized photon beam will be the first attempt to
study vector polarized $\gamma $-deuteron Compton scattering.
Experiments using tensor polarized targets are also technically
feasible. Existing
technologies like spin-exchange optical pumping \cite{op}, which will
be applied in the BLAST polarized $^1$H and $^2$H laser driven sources,
and free electron lasers can provide high quality polarized deuteron
targets and photon beams. The question is, what kind of physics can be
measured in such an experiment? Is it interesting enough to motivate the
experiments?
 
Theoretically, four vector form factors (corresponding to
the $\Delta J=1$ interaction of the deuteron field) and six tensor form factors
(corresponding
to $\Delta J=2$) are identified in~\cite{WA,Weyr2} with the lower
multipole contributions calculated using dispersion relations and potential models.
In this paper, we perform an analytic calculation of the differential
cross section of the tensor polarized $\gamma $-deuteron Compton scattering
cross section to first non-vanishing order in the effective field theory
expansion. At this order, the contributions to the cross section from
the $\Delta J=2$ amplitude are the interference terms
between the leading electric coupling diagrams and the subleading single
potential pion exchange diagrams or the subleading magnetic moment
coupling diagrams. Thus the pion effects
are of leading order. We can further isolate the pion 
contribution by setting the photon scattering angle to be 90$^{\circ }.$
At
this angle, only the pion term contributes at this order to the
tensor polarized differential cross section.\ This provides a clean way to
study the photon pion dynamics in the two nucleon sector. The experimental
precision required to measure these effects will also be discussed in the
following sections.

\section{Tensor Polarized $\gamma$-Deuteron Compton Scattering}

The process we will focus on is the low energy (below pion production
threshold) tensor polarized Compton scattering 
\begin{equation}
\gamma (\omega ,{\bf k})\stackrel
\leftrightarrow{d}\rightarrow \gamma
(\omega
^{^{\prime }},{\bf k}^{^{\prime }})d\quad , 
\end{equation}
where the incident photon of four momentum $(\omega ,{\bf k)}$ in the
deuteron rest frame (lab frame) scatters off the polarized deuteron target
to an outgoing photon of four momentum $(\omega ^{^{\prime }},{\bf k}%
^{^{\prime }}{\bf )}$. The polarization of photons and final state deuteron
are not detected. For convenience, we will work in the lab frame and choose
the ${\bf k}$ direction to be the (0,0,1) direction. The scattering
amplitude can be written in terms of scalar, vector and tensor form factors $%
S,V,$ and $T$ corresponding to the $\Delta J=0,1,2$ interactions of the
deuteron field,

\begin{equation}
\label{fg}{\cal M}=i\frac{\displaystyle e^2}{\displaystyle M_N}\left\{ S{\bf %
\varepsilon }_d{\bf \cdot \varepsilon }_d^{^{\prime }*}+{\bf \varepsilon }%
_{ijk}V_i{\bf \varepsilon }_{d_j}{\bf \varepsilon }_{d_k}^{^{\prime
}*}+T_{ij}\left( {\bf \varepsilon }_{d_i}{\bf \varepsilon }_{d_j}^{^{\prime
}*}+{\bf \varepsilon }_{d_j}{\bf \varepsilon }_{d_i}^{^{\prime }*}-\frac 23%
\delta _{ij}{\bf \varepsilon }_d{\bf \cdot \varepsilon }_d^{^{\prime
}*}\right) \right\} \quad , 
\end{equation}
where ${\bf \varepsilon }_d$ and ${\bf \varepsilon }_d^{^{\prime }}$ are the
polarization vectors of the initial and final state deuterons. Using the
power counting described in \cite{gamd}, the scalar form factor $S$
contributes to the amplitude starting at leading order (LO) in the effective
field theory expansion, while the tensor form factor $T$ and the vector
form factor $V$ contribute starting at next-to-leading order (NLO) and
next-next-to-leading order (NNLO) respectively. Squaring the amplitude to
form the cross section, the LO contribution to the cross section only comes
from the $\left| S\right| ^2$ term which is independent of the deuteron
target polarization, while the target polarization dependent $S$ and $T$
interference term contributes at NLO. Note that the vector form factor $V$
does not contribute to the cross section through the $S$ and $V$
interference
term since the final state deuteron polarization is not detected.

It is useful to define the tensor polarized differential cross section $%
d\sigma _2$ as a tensor combination of polarized differential cross sections
to eliminate the LO polarization independent effect and make the $S$ and $T$
contribution become leading. $d\sigma _2$ is defined by 
\begin{equation}
\label{def}\frac{\displaystyle d\sigma _2}{\displaystyle d\Omega }\equiv {%
\frac{\displaystyle 1}{\displaystyle 4}}\left[ 2\frac{\displaystyle d\sigma 
}{\displaystyle d\Omega }(J_z=0)-\frac{\displaystyle d\sigma }{\displaystyle %
d\Omega }(J_z=+1)-\frac{\displaystyle d\sigma }{\displaystyle d\Omega }%
(J_z=-1)\quad \right] \ \quad , 
\end{equation}
where $J_z$ indicates the polarization of the deuteron
target. We have 
\begin{equation}
\begin{array}{c}
J_z=0\quad ,\quad 
{\bf \varepsilon }_d=(0,0,1)\quad ; \\ J_z=\pm 1\quad ,\quad {\bf %
\varepsilon }_d=%
{\displaystyle {1 \over \sqrt{2}}}
(1,\pm i,0)\quad . 
\end{array}
\end{equation}

\begin{figure}[t]
\centerline{{\epsfxsize=4.0in \epsfbox{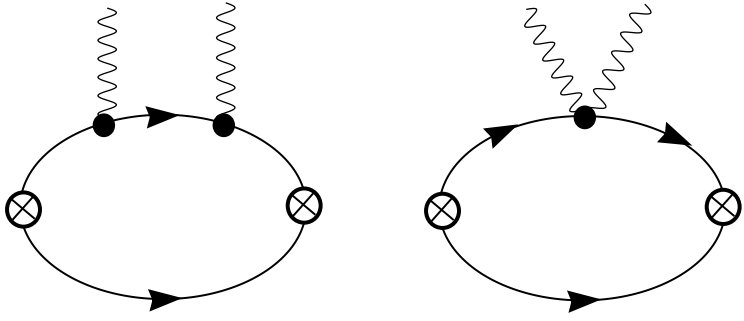}}\hskip 0.8in }
\vskip 0.1in
\noindent
\caption{\it Leading order contributions to the scalar form factor of 
$\gamma$-deuteron
  Compton scattering.
  The crossed circles denote operators that create
  or annihilate
  two nucleons with
  the quantum numbers of the deuteron. 
  The dark solid circles correspond to the photon
  coupling via
  the nucleon kinetic energy operator (minimal coupling).
The solid lines are nucleons.
  The photon crossed graphs are not shown.
}
\label{fig:gdEleclead}
\vskip .2in
\end{figure}

To predict $d\sigma _2$, we need expressions for $S$ and $T$. The  
calculation of the scalar form factor $S$ was carried out up to NLO in
\cite
{gamd}. Two distinct structures contributing to $S$ can be parameterized by
electric and magnetic form factors $F_0$ and $G_0$ as
\begin{equation}   
S=F_0{\bf \varepsilon \cdot \varepsilon }^{^{\prime }*}+G_0\
(\widehat{{\bf k%
}}\times {\bf \varepsilon \ )\cdot (}\widehat{{\bf k}}^{\prime }{\bf
\times
\varepsilon }^{^{\prime }*})\quad ,
\end{equation}
where ${\bf \varepsilon }$ and ${\bf \varepsilon }^{^{\prime }}$ are the
polarization vectors of the initial and final state photons, $\widehat{{\bf k%
}}$ and $\widehat{{\bf k}}^{\prime }$ are unit vectors in the direction of 
${\bf k}$ and ${\bf k}^{\prime}$. At LO (denoted by a superscript on the
form
factors), $%
F_0$ receives contributions from the electric coupling of the $N^{\dagger }%
{\bf D}^2N$ operator (minimal coupling), as shown in Fig.~
\ref{fig:gdEleclead}, to be 
\begin{eqnarray}
F_0^{LO}=&-&\left[ 
\frac{\displaystyle \sqrt{2}\gamma }{\displaystyle \left| \omega \right| 
\sqrt{1-\cos\theta}}\tan ^{-1}\left( \frac{\displaystyle \left| \omega
\right| \sqrt{1-\cos\theta}}{\displaystyle 2\sqrt{2}\gamma }\right) +%
\frac{\displaystyle 2\gamma ^4}{\displaystyle 3M_N^2\omega ^2}-\frac{\displaystyle 
2\gamma (\gamma ^2-M_N\omega -i\epsilon )^{3/2}}{\displaystyle 
3M_N^2\omega ^2}\right] \nonumber \\ &+&\left[\omega \rightarrow -\omega
\right]\quad ,
\label{f0}
\end{eqnarray}
where cos~$\theta =\widehat{{\bf k}}\cdot \widehat{{\bf k}}^{\prime }$ is
the
cosine of the angle between the incident and outgoing photons and $\gamma =%
\sqrt{M_NB}$ is the deuteron binding momentum. In eq.~(\ref{f0}) terms
suppressed by additional factors of ${\bf k}^2/(m_N\omega)$ (i.e., recoil
effects) are neglected since they only contribute to NNLO differential
cross sections. At LO, the magnetic contribution vanishes, 
\begin{equation}
G_0^{LO}=0\quad . 
\end{equation}
We do not show the NLO results for $F_0$ and $G_0$ here since we only
calculate $d\sigma _2$ to its first non-vanishing order.

The tensor form factor $T$ has more distinct structures. However, up
to NLO, after neglecting pion contributions suppressed by
factors of ${\bf k^2}/m_\pi^2$, $T$ can be  
parameterized by electric and magnetic form factors $F_2$ and $G_2$ as

\begin{equation}
T_{ij}=\ F_2{\bf \varepsilon }_i{\bf \varepsilon }_j^{^{\prime }*}+G_2\ (%
\widehat{{\bf k}}\times {\bf \varepsilon \ )}_i (\widehat{{\bf
k}}^{^{\prime }}{\bf \times
\varepsilon }^{^{\prime }*})_j\quad . 
\end{equation}
$F_2$ and $G_2$ are related to the form factors $P2(E1,E1)$ and
$P2(M1,M1)$ defined in \cite{WA,Weyr2} by a normalization factor. 
At LO, $F_2$ and $G_2$ vanish, 
\begin{equation}
F_2^{LO}=0\quad ,\quad G_2^{LO}=0\quad . 
\end{equation}


\begin{figure}[t]
\centerline{\epsfxsize=4.0in \epsfbox{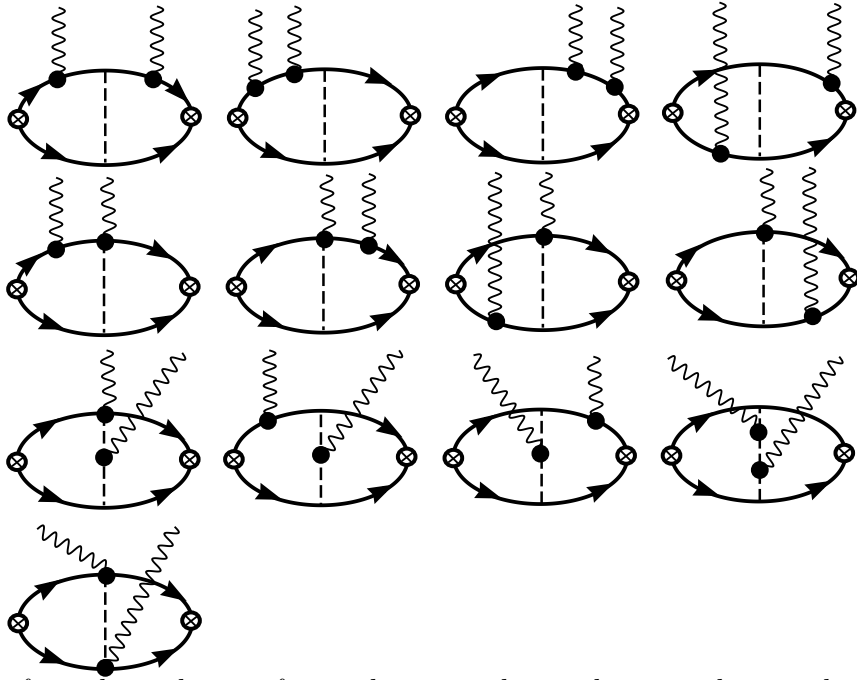}}
\noindent
\caption{\it Graphs from the exchange of a single
  potential pion that
  contribute to the tensor form factor of $\gamma$-deuteron Compton
scattering
  at NLO.
  The crossed circles denote operators that create or
  annihilate two nucleons with
  the quantum numbers of the deuteron.
  The solid circles correspond to the photon coupling via
  the nucleon or meson kinetic energy operator
  (minimal coupling)
  or from the gauged axial coupling to the meson field.
  Dashed lines
  are mesons and solid lines are nucleons.
  Photon crossed graphs are not shown.   
  }
\label{fig:Polsubpi}
\vskip .2in
\end{figure}

At NLO $F_2$ receives contributions from the electric coupling of the one
potential pion exchange diagrams shown in Fig.~\ref{fig:Polsubpi} and
gives
a
renormalization scale $\mu $ independent contribution 
\begin{eqnarray}
   F_{2}^{NLO}  & = &
 {g_A^2\gamma  \over 240\pi f^2 M_N\omega ^2}
 \left\{
   \left[ m_\pi ^2(17m_\pi ^2-28\gamma ^2)-24M_N^2\omega ^2\right] \log
\left[ 
{m_\pi +2\gamma  \over \mu}
\right]
\right.\nonumber\\ & & \left.
    \ -\    m_\pi ^2(m_\pi ^2+4\gamma ^2-4M_N\omega)
\log
\left[ 
{\displaystyle {m_\pi +2\sqrt{\gamma ^2-M_N\omega -i\epsilon } \over \mu}}
\right] 
\right.\nonumber\\ & & \left.
  \ -\  8\left[ m_\pi ^2(2m_\pi ^2-4\gamma ^2+M_N\omega )-3M_N^2\omega
^2\right] \log \left[ 
{\displaystyle {m_\pi +\gamma +\sqrt{\gamma ^2-M_N\omega -i\epsilon } \over \mu}}
\right]
\right.\nonumber\\ & & \left.
\ +\   {32M_N\omega \left[ 2m_\pi \gamma ^2-M_N\omega (m_\pi -\gamma )\right] 
\over m_\pi +\gamma +\sqrt{\gamma ^2-M_N\omega -i\epsilon }}
\right.\nonumber\\ & & \left.
\ +\ {\displaystyle {4M_N^2\omega ^2 ( 5m_\pi ^2-12\gamma ^2) 
 \over (m_\pi +2\gamma )^2\ }}
\right.\nonumber\\ & & \left.
\ -\ 2\left[ m_\pi ^3+4\left( 2m_\pi ^2-M_N\omega \right) (m_\pi -\gamma
)\right] \left( \gamma -\sqrt{\gamma ^2-M_N\omega -i\epsilon }\right)
\right\}\nonumber\\ & & \left.
\ +\  \left\{\omega \rightarrow -\omega \right\}\right.
\ \ \ ,
\label{fpi}
\end{eqnarray}
where $f=132$ MeV is the pion decay constant.
In the computation of the pion diagrams, the pion  
propagators with photon momentum dependence have the form
\begin{equation}
{1\over -(p_0-\omega)^2+({\bf p-k})^2+m_\pi^2}\approx
{1\over{\bf p}^2-2 {\bf p}\cdot{\bf k}+m_\pi^2}\quad ,
\end{equation}
where $(p_0,{\bf p})$ is the loop momentum. $p_0$ is of higher
order in the power counting compared with the other scales in the
propagators and can be neglected. We keep the first term in the
${\bf k}$ expansion of the pion propagators to get the result of 
eq.~(\ref{fpi}). The error of this approximation is of order  
${\bf k}^2/m_\pi ^2$ of eq.~(\ref{fpi}) and of even higher order in
powers of ${\bf k}^2/m_\pi ^2$ at $\theta=0$ and $\theta=\pi$.
Because the error is formally of NLO in power counting,
strictly speaking we have not presented the complete calculation
at NLO. However, numerically the terms omitted are NNLO.

\begin{figure}[t]
\centerline{{\epsfxsize=1.2in \epsfbox{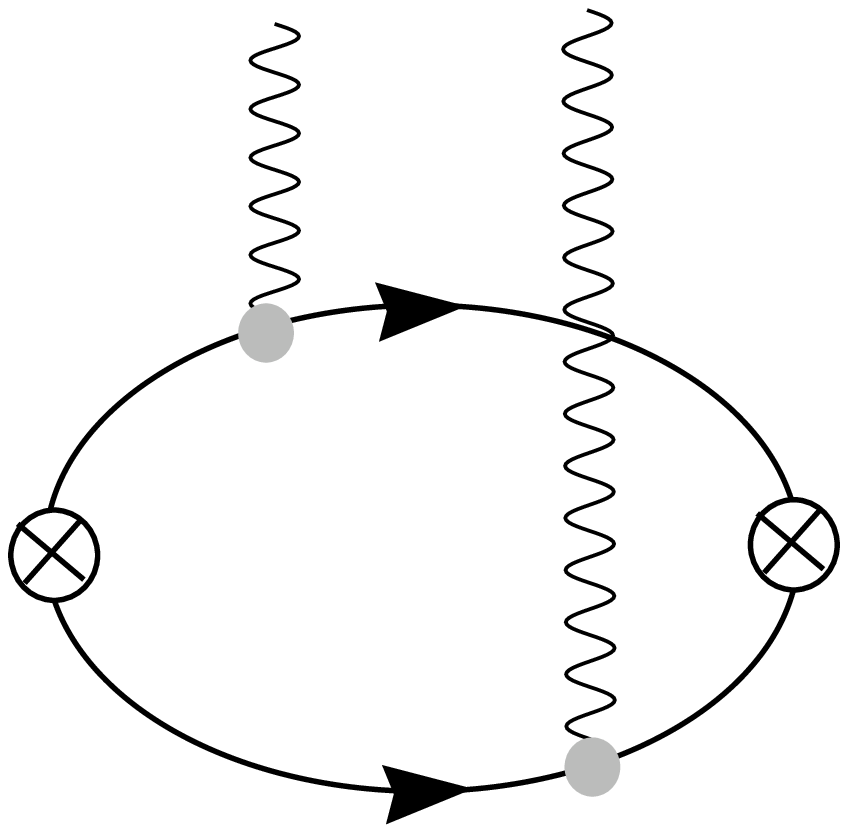}} }
\vskip 0.1in
\centerline{ {\epsfxsize=5in \epsfbox{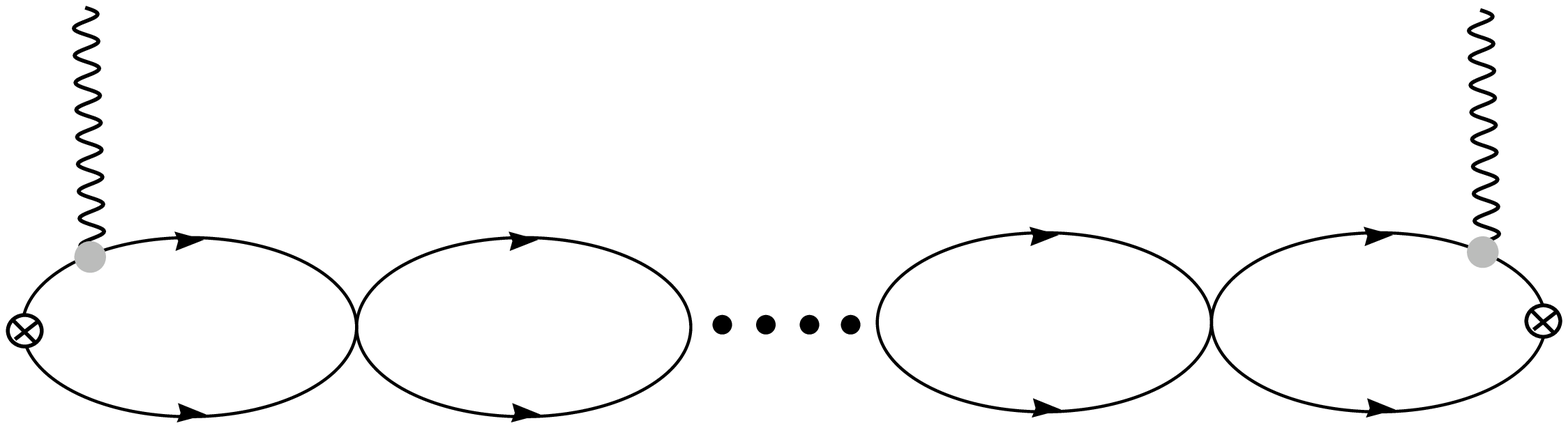}}}
\noindent
\caption{\it Graphs from insertions of the nucleon
  magnetic moment interaction that contribute to the
tensor form factor of $\gamma$-deuteron
  Compton scattering at NLO.
  The crossed circles denote operators that create or
  annihilate
  two nucleons with
  the quantum numbers of the deuteron.
The light solid circles denote the nucleon magnetic
moment operator.
The solid lines are nucleons.
  The bubble chain arises from insertions of the four
  nucleon operator
  with coefficient $C_0^{(\si)}$ or $C_0^{(\siii)}$ .
}
\label{fig:Maglead}
\vskip .2in
\end{figure}

The magnetic moment coupling diagrams shown in Fig.~\ref{fig:Maglead}
contribute to $G_2$
at NLO as 
\begin{eqnarray}
G_2^{NLO} 
&=&- 
\frac{\displaystyle  2(\kappa ^{(0)^2}-\kappa ^{(1)^2})\gamma (\gamma
-\sqrt{\gamma
^2-M_N\omega -i\epsilon })}{\displaystyle  M_N^2}\nonumber \\ 
&+&\frac{\displaystyle  \kappa ^{(1)^2}\gamma (\gamma -\sqrt{\gamma
^2-M_N\omega -i\epsilon })^2{\cal A}_{-1}^{(^1S_0)}(\omega -B)}{\displaystyle 2\pi M_N}\nonumber \\ 
&-& \frac{\displaystyle \kappa ^{(0)^2}\gamma (\gamma -\sqrt{\gamma
^2-M_N\omega -i\epsilon })^2{\cal A}_{-1}^{(^3S_1)}(\omega
-B)}{\displaystyle 2\pi M_N}\nonumber \\
&+&(\omega \rightarrow
-\omega )\quad  ,
\end{eqnarray}
where $\kappa ^{(0)}=(\kappa_p+\kappa_n)/2$ and $\kappa
^{(1)}=(\kappa_p-\kappa_n)/2$ are isoscalar and isovector nucleon magnetic
moments in nuclear magnetons, with $\kappa_p=2.79285$ and
$\kappa_n=-1.91304$.
The leading order nucleon nucleon scattering amplitudes contribute to the
magnetic moment diagrams through the $C_0$ bubble chains and has the
expression \cite{KSW,KSW2}

\begin{equation}
{\cal A}_{-1}^{(^1S_0),(^3S_1)}(\omega -B)=\frac{-\displaystyle %
C_0^{(^1S_0),(^3S_1)}}{\displaystyle 1+\displaystyle
C_0^{(^1S_0),(^3S_1)}\frac{\displaystyle M_N}{\displaystyle 4\pi }(\mu -%
\sqrt{-M_N(\omega -B)-i\varepsilon }\ )}\quad . 
\end{equation}
The renormalization scale $\mu $ dependence in the denominator is canceled
by the $\mu $ dependence of $C_0^{(\si)}$ and $C_0^{(^3S_1)}$, as
required.
Values of $C_0^{(\si)}$ and $C_0^{(^3S_1)}$ have been determined from
nucleon nucleon scattering in the $\si$ and $^3S_1$ channels
\cite{KSW,KSW2}
to be $C_0^{(\si)}=-3.34\ {\rm fm^2}$ and $C_0^{(^3S_1)}=-5.51\ {\rm
fm^2}$
at 
$\mu =m_\pi .$

Having obtained the leading non-vanishing contributions for $S$ and $T$, we
now give the leading non-vanishing expressions for $d\sigma _2$. At LO, $%
d\sigma _2$ vanishes, 
\begin{equation}
\frac{\displaystyle d\sigma _2^{LO}}{\displaystyle d\Omega }=0\quad . 
\end{equation}
While at NLO,

\begin{equation}
\label{sig2}\frac{\displaystyle d\sigma _2^{NLO}}{\displaystyle d\Omega }=%
\frac{\displaystyle \alpha ^2}{\displaystyle 2M_N^2}\left[
Re[F_0^{LO}F_2^{NLO^{*}}](1+\cos ^2\theta )+2Re[F_0^{LO}G_2^{NLO^{*}}]\cos
\theta \right] \quad . 
\end{equation}

\begin{figure}[t]
\centerline{\epsfxsize=3.7in \epsfbox{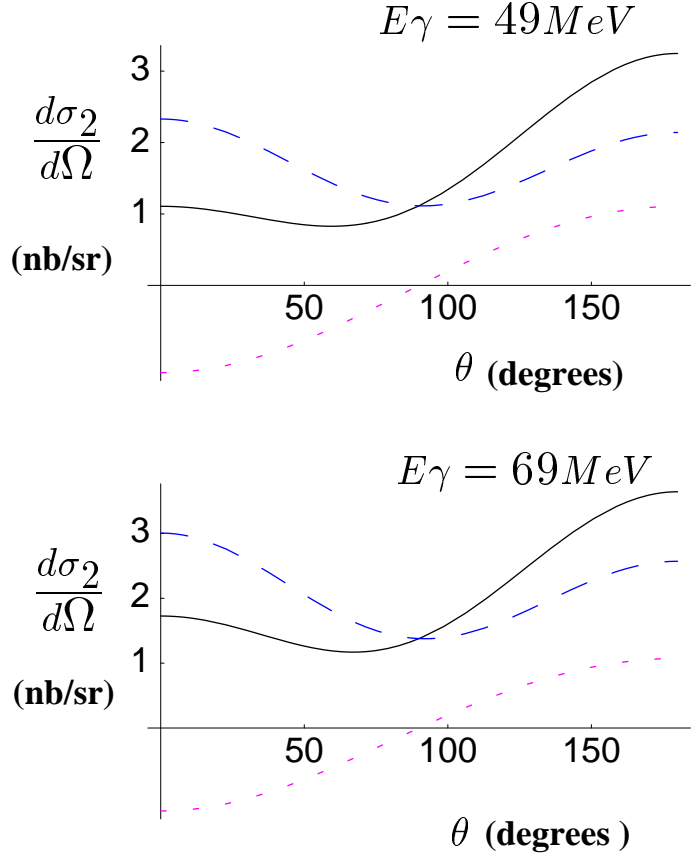}}
\noindent
\caption{\it The differential cross section for
  tensor polarized $\gamma$-deuteron Compton scattering to first
  non-vanishing order at incident photon
  energies of
  $E_{\gamma}=49\ {\rm MeV}$  and  $69\  {\rm MeV}$.
  The dashed curves correspond to the pion interference term 
  contributions. The dotted curves correspond to the magnetic moment
interference term
  contributions. The solid curves correspond to the sum of the pion and 
  magnetic moment interference terms.}
\label{fig:dsig24969}
\vskip .2in
\end{figure}

There are two terms in $d\sigma _2^{NLO}.$ The first term comes from the
interference between the LO electric coupling diagrams and the NLO one
potential
pion exchange diagrams. The second term comes from the interference 
between the LO electric coupling
diagrams and the NLO magnetic moment coupling diagrams. 
In Fig.~\ref{fig:dsig24969}, the one
pion interference term contributing to $d\sigma _2\ $ are plotted as
the
dashed curves at incident photon energies of 49 and 69 MeV. The angular
distributions are dominated by the $(1+\cos ^2\theta)$ factor that comes
from the sum of the inner product squares of the incident and
outgoing electric fields over the two different modes (electric
fields parallel or perpendicular to the scattering plane). The small
asymmetry comes from the finite size of the deuteron. The magnetic moment
interference terms are plotted as the dotted curves. The $\cos\theta$
factor comes from the sum of the inner product of the incoming and
outgoing electric fields times the inner product of the incoming and
outgoing magnetic fields over
the two different modes. The sum of the two terms that form the NLO
contributions to $d\sigma _2/d\Omega $ are plotted as solid lines. The
result is dominated by the pion interference terms. Higher order
corrections could be as large as 30\% of the contributions we show in this
figure.


\begin{figure}[t]
\centerline{\epsfxsize=3.7in \epsfbox{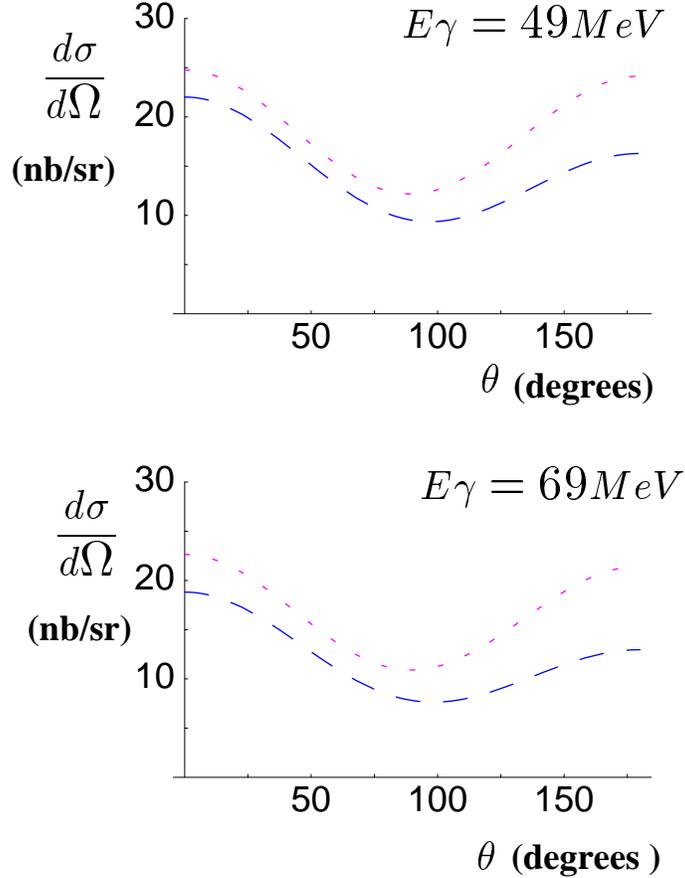}}
\noindent
\caption{\it The differential cross section for
  polarized $\gamma$-deuteron Compton scattering at incident photon
  energies of
  $E_{\gamma}=49\ {\rm MeV}$  and  $69\  {\rm MeV}$.
  The dotted curves correspond to the NLO deuteron $J_z=0$ result.
  The dashed curves correspond to the NLO deuteron $J_z=1$ result.
  }
\label{fig:polrd4969}
\vskip .2in
\end{figure}
%
   
\begin{figure}[t]
\centerline{\epsfxsize=3.7in \epsfbox{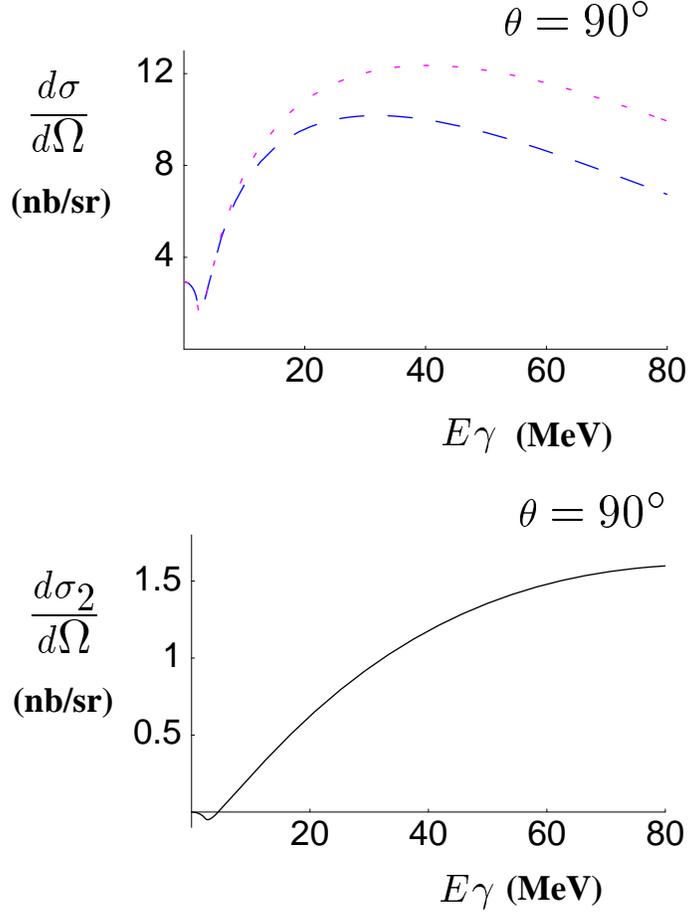}}
\noindent
\caption{\it The differential cross section for
  polarized $\gamma$-deuteron Compton scattering at $90^{\circ }$.
  The dotted curve corresponds to the NLO deuteron $J_z=0$ result. 
  The dashed curves corresponds to the NLO deuteron $J_z=1$ result.
  The solid curve corresponds to the tensor polarized result which is
  half of the difference between the dotted and dashed curves.
  }
\label{fig:90deg4969}
\vskip .2in
\end{figure}
%


\begin{figure}[t]
\centerline{\epsfxsize=4.0 in \epsfbox{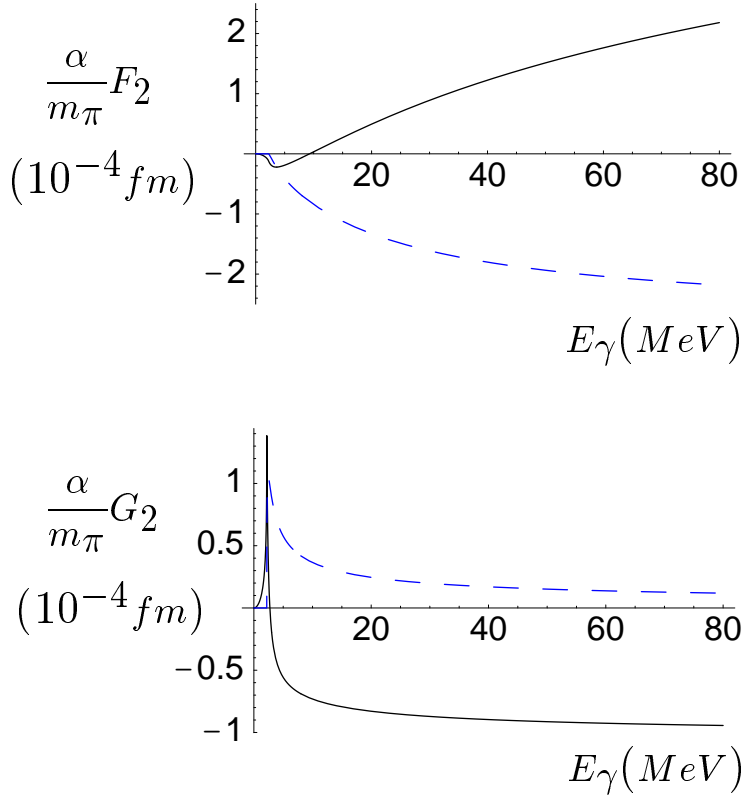}}
\noindent
\caption{\it Form factors $F_2$ and $G_2$ calculated to the first
  non-vanishing order (NLO) shown as
  functions of photon energy. The solid lines are real parts and the
  dashed lines are imaginary parts. 
  }
\label{fig:f2g2}
\vskip .2in
\end{figure}   

To indicate the relative size of $d\sigma _2,$ we write the
$J_z=0$ and $J_z=\pm 1$ differential
cross sections in terms of the unpolarized and
tensor polarized differential cross sections, $d\sigma _0$ and $d\sigma _2$,
as 
\begin{eqnarray}
\frac{\displaystyle d\sigma }{\displaystyle d\Omega
}(J_z=0)&=&\frac{\displaystyle d\sigma _0^{}}{\displaystyle d\Omega
}+%
{4 \over 3}\frac{\displaystyle d\sigma _2^{}}{\displaystyle d\Omega
}\quad ,\nonumber \\ \frac{\displaystyle d\sigma }{\displaystyle d\Omega }(J_z=\ \pm 1)&=&\frac{\displaystyle d\sigma
_0^{}}{\displaystyle d\Omega }-{2 \over 3}
\frac{\displaystyle d\sigma _2^{}}{\displaystyle d\Omega }\quad .
\label{poldsig}
\end{eqnarray}
The unpolarized differential cross section,

\begin{equation}
{\frac{\displaystyle d\sigma _0}{\displaystyle d\Omega }}=\frac{%
\displaystyle \alpha ^2}{\displaystyle 2M_N^2}\left[ \ \left( \left|
F_0\right| ^2+\left| G_0\right| ^2\right) \ (1+\cos ^2 \theta )+4Re\left[
F_0G_0^{*}\right] \cos \theta \right] \quad ,
\end{equation}
up to NLO can be found in \cite{gamd}. Note that we have used 
$d\sigma ($ $J_z=+1)$ =$d\sigma ($ $J_z=-1)$ in
eq.~(\ref{poldsig}) as photons are not circularly
polarized. This is a consequence of parity conservation. In
Fig.~\ref{fig:polrd4969}, we plot the $J_z=0$
and $J_z=\pm 1$ differential
cross sections as dotted and dashed curves respectively at photon energies
of 49 and 69 MeV. The difference between each set of dotted and dashed
curves gives $2d\sigma _2/d\Omega $ while the weighted sum
gives $d\sigma_0/d\Omega.$ At both energies, 
$d\sigma _2/d\Omega $ at backward angles
can
be extracted if the error bars in $J_z=0$ and $J_z=\pm 1$ differential cross
sections are $\sim 10 \%$, as that of the unpolarized experiment 
\cite{Lucas}
performed four years ago. With error bars of 4\%, the forward angle part
of $d\sigma_2/d\Omega$ would also be measurable. In these estimates, we have 
assumed that the higher order corrections
to $d\sigma_2/d\Omega $ are 30\%.

We can isolate the one pion interference term from $d\sigma
_2^{NLO}$ by going to $\theta =90^{\circ }$. As can easily be
seen from eq.~(\ref{sig2}), at $\theta =90^{\circ }$ the magnetic
interference term vanishes. This
provides a clean way to study the photon pion dynamics in the two nucleon
sector. In Fig.~\ref{fig:90deg4969}, the NLO differential cross
sections at $\theta =90^{\circ }$ for deuteron polarizations $J_z=0$ and 
$J_z=\pm 1$ are plotted as the dotted and dashed curves while 
$d\sigma _2/d\Omega $ is 
plotted as the solid curve. Again, with measurements of the $%
J_z=0$ and $\pm 1$ polarization made to within 7\%, the
pion interference
term effects would be observable for photon energies between 40 and 80 MeV
with the higher order effect estimated to be 30\%. Note that the effective
field theory expansion power counting is expected to break down at photon
energy of about
100 MeV (recent works by Mehen and Stewart \cite{MeStew} suggest that in
fact the breakdown scale may be much higher). But the neglected NLO terms
which are
suppressed by factors of ${\bf k^2}/m_\pi^2$ could contribute 
a $30\%$ effect at 80 MeV. Thus the
80 MeV photon energy sets an upper bound for the
calculations performed in this paper.

It is also interesting to compare our results with those from potential
model calculations \cite{WA,Weyr2}. In Fig.~\ref{fig:f2g2}, we show 
the fist non-vanishong order (NLO) results of $F_2$ and $G_2$
multiplied by a constant to compare with the
form factors $P2(E1,E1)$ and $P2(M1,M1)$ calculated and plotted in
\cite{WA}. The agreement on $G_2$ is $\sim 10\%$ while the agreement on
$F_2$ is $\sim 30$-$40\%$.

\section{Conclusions}

We have presented analytic expressions for the differential cross section of
tensor polarized $\gamma $-deuteron Compton scattering in an effective
field theory expansion. The first
non-vanishing contributions are the interference terms between the leading
electric coupling diagrams and the subleading single potential pion
exchange diagrams, and between the leading
electric coupling diagrams and the
subleading magnetic moment coupling diagrams. At 90$^{\circ }$ photon
scattering angle, only the pion term contributes at this order to
the tensor polarized differential cross section.\ This provides a clean way
to study the photon pion dynamics in the two nucleon sector. For photon
energy between 40 and 80 MeV, the one pion interference term contributions
would be measurable at 90$^{\circ }$ provided the uncertainty in the
$\gamma $-deuteron Compton scattering experiments are $\lsim 7\%$. At
backward angles, the magnetic interference term adds to
the pion term to give an effect which could be seen if the
uncertainty in the measurements are $\lsim 10\%$. 

\vskip 0.5in

The author would like to thank Martin Savage, Roxanne Springer, Harald
Griesshammer, Jerry Miller, Jon Karakowski ,and Daniel Phillips for
helpful discussions and Hartmuth Arenhovel for correspondence. This
work is supported in part by the U.S. Dept. of Energy
under Grants No. DE-FG03-97ER41014.

\end{document}